# Low frequency resistive instability in a dusty Hall thruster plasma

Jasvendra Tyagi[1], Hitendra K. Malik[1*], Yves Elskens[2], Nicolas Lemoine[3], and Fabrice Doveil[2]

[1]*Plasma Waves and Particle Acceleration Laboratory, Department of Physics, Indian Institute of Technology Delhi, New Delhi – 110 016, India*

[2]*UMR 7345 CNRS - Aix-Marseille Université, PIIM, case 322, Turbulence plasma, av. esc. Normandie-Niemen, 13397 Marseille cedex 20, France*

[3]*Institut Jean Lamour IJL, UMR 7198 CNRS, Université de Lorraine, BP 70239, 54506 Vandoeuvre-lès-Nancy, France*

E-mail: *h.k.malik@hotmail.com, [1]jasvendratyagi83@gmail.com, [2]yves.elskens@univ-amu.fr, [3]nicolas.lemoine@univ-lorraine.fr, [2]fabrice.doveil@univ-amu.fr

**ABSTRACT**

Along with crossed electric and magnetic fields in a Hall thruster, a radial component of electric field is generated that takes ions toward the walls, which causes sputtering and produces dust contamination in the thruster plasma. Considering negatively charged dust particles in the Hall thruster, we approach analytically the resistive instability by taking into account the oscillations of dust particles, ions and electrons along with finite temperatures of ions and electrons. In typical Hall thruster regimes, the resistive instability growth rate increases with higher collision rates in the plasma, stronger magnetic field but it decreases with higher mass of the dust and higher temperature of the ions and electrons. In comparison with dust-free models, the presence of dust results into a drop of the resistive instability growth rate by three orders of magnitude, but the growth rate increases slowly for dust densities within the typical range.

# I. INTRODUCTION

After about a century of development[1], recent years witnessed the successful use of electric propulsion in space missions.[2] Among electric propulsion devices, the Hall thruster stands among the most promising ones for small satellites because of its high efficiency and thrust density. Improving the efficiency of the Hall thruster requires a detailed knowledge of physical processes occurring in the plasma of a thruster.[3] Since the plasma in Hall thrusters is far from equilibrium and is placed in a rather strong magnetic field, both features create conditions for plasma instabilities to occur. These instabilities influence the operation of the thruster. Therefore, the instabilities in the Hall thrusters have been and remain a most important research topic.[4-20]

Various oscillations with different length and time scales along with several kinds of physical phenomena occur in the magnetized and inhomogeneous plasma of a Hall thruster from a few tens of kHz up to tens of MHz, from millimeter scale[7] to centimeter scale[8]. Instabilities at centimetre scale from a few MHz to tens of MHz, whose phase velocity is close to the azimuthal electron drift velocity and whose wavevector is almost azimuthal with a small axial component, have been extensively studied experimentally, for example by Lazurenko and Albarède[8] and by Litvak *et al.*[19]. Several mechanisms have been considered to explain them. Inhomogeneity in the electron flow further leads to Rayleigh type instability.[20-24] Also, plasma resistivity has been found to induce resistive instabilities corresponding to lower hybrid waves and Alfven waves, as well as the resistive instabilities associated with the axial ion flows.[14, 26] Nejoh's group has done a tremendous work concerning collisional sheath structure and fluctuations in Hall thruster plasmas including charging near the ion engine.[9-13] The plasma perturbations in the acceleration channel of a Hall current plasma discharge have been found to be unstable due to the resistive coupling to the electron's $\vec{E} \times \vec{B}$ drift in the presence of collisions.[14] Singh and Malik investigated the resistive instability of electrostatic and electromagnetic waves under the presence of collisions[15] : high frequency electromagnetic resistive instabilities are found to occur in a Hall thruster plasma in the presence of collisions and ionization.[16] Chesta *et al.*[17] estimated theoretically the growth rate and frequencies of predominantly axial and azimuthally propagating plasma disturbances.

In most of these studies, dust particles are neglected, though they become an important plasma species in the Hall thrusters due to the sputtering and some impurity level in the propellant. These dust grains are much heavier than the ions, and get charged by the impact of the electrons and ions. They may collect either negative or positive charge depending upon their charging process. It is observed that the current carried by the highly mobile electrons plays the dominant role and, as a result, the dust grains are negatively charged in a plasma.[25] The negatively charged dust particles will hinder the movement of ions, and hence the thrust and efficiency of the thruster will be drastically lowered. Moreover, these particles will collect large number of electrons, lowering the ionization in the plasma.

In the scope of this article, we focus on the effect of dust on the resistive instability caused by the resistive coupling of electrons flow with the oscillations in the presence of collisions. As opposed to some earlier cases, we consider the oscillations of dust particles in addition to the movement of ions and electrons. In order to uncover the role of dust particles in this instability, we compare the growth rate of the instability for both the cases of presence and absence of dust particles in the acceleration chamber of the thruster.

Section II of the manuscript presents the theoretical model of the resistive instability that includes the derivation of dispersion equation. Section III is devoted to the limiting cases of the calculations. Results are discussed in Section IV, and the conclusions are given in the final Section V.

**II. THEORETICAL MODEL OF RESISTIVE INSTABILITY**

As mentioned above, we consider a Hall thruster with plasma consisting of ions, electrons and negatively charged dust particles (of same size) immersed in a magnetic field $\vec{B} = B\,\hat{z}$ such that electrons are magnetized while ions and dust particles are unmagnetized. Because of the axial electric field $\vec{E}$ (along $x$ - axis), the electrons have an $\vec{E} \times \vec{B}$ drift in the azimuthal direction ($y$-axis) whereas the movement of ions and dust particles is restricted along $x$-axis. Unlike most previous works, we take into account the motion of all the species, i.e. dust (density $n_d$, mass $m_d$, velocity $\vec{v}_d$, temperature $T_d$), ion (density $n_i$, mass $m_i$, velocity $\vec{v}_i$, temperature $T_i$), and electrons (density $n_e$, mass $m_e$, velocity $\vec{v}_e$, temperature $T_e$) for the excitation of waves and instability. The basic fluid equations for all the species read

$$\frac{\partial n_j}{\partial t} + \vec{\nabla} \cdot (\vec{v}_j n_j) = 0, \tag{1}$$

$$m_j n_j \left[ \frac{\partial \vec{v}_j}{\partial t} + (\vec{v}_j \cdot \vec{\nabla})\vec{v}_j \right] = s_j n_j \vec{E} + s_j n_j (\vec{v}_j \times \vec{B}) - c_j n_j m_j \nu \vec{v}_j - \vec{\nabla} p_j. \tag{2}$$

where $j$ = e, i, d. In Eq. (2), $s_e = -e$ for electrons, $s_d = -Z_d e$ for dust and $s_i = +e$ for ions.

Since dust particles and ions are slower than electrons, they are unmagnetized and the second term of the right hand side of Eq. (2) vanishes for these species. Collisions with neutrals are modeled by the third term on the right hand side of Eq. (2) for electrons with $v$ as their collision frequency. As we do not consider collision between other particles, we set $c_e = 1$ and $c_i = c_d = 0$. Finally, the pressure oscillations are modeled with the adiabatic approximation $p_j / n_j^{Y_j}$ = constant, where $Y_e = Y_i = 2$ as we consider all three species as monatomic fluids moving in two space dimensions, normal to the magnetic field. The pressure gradient term for the dust particles is neglected in view of their very low temperature ($T_d = 0$).

Since the ions are unmagnetized and accelerated along the axial direction of the chamber, we consider their drift only in the $x$ – direction and neglect their motion in azimuthal and radial directions ($\vec{v}_{i0} = v_{i0}\, \hat{x}$). Moreover, in view of their heavy mass the motion (drift) of the dust particles is neglected, i.e. $v_{d0} = 0$. Since electrons are magnetized, we consider their $\vec{E} \times \vec{B}$ drift in the $y$ – direction and neglect $x$ and $z$ components ($\vec{v}_{e0} = v_{e0}\, \hat{y}$). The linearized form of Eqs. (1) and (2) for all the species yields

$$\frac{\partial n_{i1}}{\partial t} + v_{i0} \frac{\partial n_{i1}}{\partial x} + n_{i0} (\vec{\nabla} \cdot \vec{v}_{i1}) = 0, \tag{3}$$

$$\frac{\partial \vec{v}_{i1}}{\partial t} + v_{i0} \frac{\partial \vec{v}_{i1}}{\partial x} = \frac{e\vec{E}_1}{m_i} - \frac{\vec{\nabla} p_i}{m_i n_{i0}}, \tag{4}$$

$$\frac{\partial n_{e1}}{\partial t} + v_{e0} \frac{\partial n_{e1}}{\partial y} + n_{e0} (\vec{\nabla} \cdot \vec{v}_{e1}) = 0, \tag{5}$$

$$\frac{\partial \vec{v}_{e1}}{\partial t} + v_{e0} \frac{\partial \vec{v}_{e1}}{\partial y} = -\frac{e}{m_e} (\vec{E}_1 + \vec{v}_{e1} \times \vec{B}) - v \vec{v}_{e1} - \frac{\vec{\nabla} p_e}{m_e n_{e0}}, \tag{6}$$

$$\frac{\partial n_{d1}}{\partial t} + n_{d0} (\vec{\nabla} \cdot \vec{v}_{d1}) = 0, \tag{7}$$

$$\frac{\partial \vec{v}_{d1}}{\partial t} = -\frac{eZ_d \vec{E}_1}{m_d}. \tag{8}$$

In order to investigate the oscillations through the solution of the above equations, we set $A_1 = A \exp(i\omega t - i\vec{k} \cdot \vec{r})$ for the first order quantities ($n_{i1}$, $n_{e1}$, $n_{d1}$, $\vec{v}_{i1}$, $\vec{v}_{e1}$, $\vec{v}_{d1}$, $\vec{E}_1$). Here $\omega$ is the frequency of oscillations and $\vec{k}$ is the wave propagation vector taken in the ($x$, $y$) plane in view of the fact that an azimuthally propagating wave gets tilted in real situation and develops an axial component of its wave vector. In view of large density gradient scale length $L$ or the weak inhomogeneity, we use the condition $\upsilon_i \ll \omega L$.

For ion and electron species, the pressure oscillations are obtained from density oscillations as $p_{j1} = Y_j T_{j0} n_{j1}$. Defining $V_{thj} = \sqrt{Y_j T_{j0}/m_j}$ (viz. $Y_j^{1/2}$ times the thermal velocity for species $j$), pressure oscillations are given by $p_{j1} = m_j V_{thj}^2 n_{j1}$.

From the equation of continuity and the equation of motion of the ions, we then obtain the expression

$$n_{i1} = \frac{ek^2 n_{i0}\phi_1}{m_i\left[(\omega - k_x \upsilon_{i0})^2 - k^2 V_{thi}^2\right]}, \tag{9}$$

where $k^2 = k_x^2 + k_y^2$.

Similarly, we get for the electrons

$$n_{e1} = \frac{n_{e0}}{(\omega - k_y \upsilon_{e0})}(k_x \upsilon_{e1x} + k_y \upsilon_{e1y}), \tag{10}$$

and, on introducing $\hat{\omega} = \omega - k_y \upsilon_{e0} - i\nu$, their equation of motion reads

$$\hat{\omega}\upsilon_{e1x} = i\Omega \upsilon_{e1y} - \frac{e}{m_e} k_x \phi_1 + V_{the}^2 k_x \frac{n_{e1}}{n_{e0}},$$

$$\hat{\omega}\upsilon_{e1y} = -i\Omega \upsilon_{e1x} - \frac{e}{m_e} k_y \phi_1 + V_{the}^2 k_y \frac{n_{e1}}{n_{e0}},$$

with the electron cyclotron frequency $\Omega = \frac{eB}{m_e}$. Noting that

$$\hat{\omega}(k_x \upsilon_{e1y} - k_y \upsilon_{e1x}) = -i\Omega(k_x \upsilon_{e1x} + k_y \upsilon_{e1y}),$$

these equations imply

$$(\Omega^2 - \hat{\omega}^2)\vec{k}\cdot\vec{\upsilon}_{e1} = \hat{\omega} k^2 \left\{\frac{e}{m_e}\phi_1 - V_{the}^2 \frac{n_{e1}}{n_{e0}}\right\}.$$

Substituting the latter expression in Eq. (10), and noting that magnetic fields in Hall thrusters are large enough for taking $\Omega \gg \max(\omega, k_y \upsilon_{e0}, \nu)$, yields the electron density in terms of the electric potential oscillations $\phi_1$ in the form

$$n_{e1} = \frac{en_{e0}\hat{\omega} k^2 \phi_1}{m_e\left[\Omega^2(\omega - k_y \upsilon_{e0}) + \hat{\omega} k^2 V_{the}^2\right]}. \tag{11}$$

Finally, the dust density oscillations are given by

$$n_{d1} = -\frac{ek^2 Z_d n_{d0} \phi_1}{m_d \omega^2}. \tag{12}$$

Using the plasma frequencies $\omega_{pe} = \sqrt{\frac{e^2 n_{e0}}{m_e \varepsilon_0}}$, $\omega_{pi} = \sqrt{\frac{e^2 n_{i0}}{m_i \varepsilon_0}}$ and $\omega_{pd} = \sqrt{\frac{Z_d^2 e^2 n_{d0}}{m_d \varepsilon_0}}$, the linearized form of Poisson's equation $\varepsilon_0 \nabla^2 \phi_1 = e(n_{e1} - n_{i1} + Z_d n_{d1})$ reduces to

$$-k^2 \phi_1 = \frac{\omega_{pe}^2 \hat{\omega} k^2 \phi_1}{\Omega^2(\omega - k_y \upsilon_{e0}) + \hat{\omega} k^2 V_{the}^2} - \frac{\omega_{pi}^2 k^2 \phi_1}{(\omega - k_x \upsilon_{i0})^2 - k^2 V_{thi}^2} - \frac{\omega_{pd}^2 k^2 \phi_1}{\omega^2}. \tag{13}$$

Since the perturbed potential $\phi_1 \neq 0$, we have from Eq. (13)

$$\frac{\omega_{pe}^2 \hat{\omega}}{\Omega^2(\omega - k_y \upsilon_{e0}) + \hat{\omega} k^2 V_{the}^2} + \frac{(\omega - k_x \upsilon_{i0})^2 - k^2 V_{thi}^2 - \omega_{pi}^2}{(\omega - k_x \upsilon_{i0})^2 - k^2 V_{thi}^2} - \frac{\omega_{pd}^2}{\omega^2} = 0. \tag{14}$$

On rewriting the above equation in polynomial form with respect to $\omega$ we obtain the dispersion equation

$$\omega^5(\omega_{pe}^2 + \Omega^2 + k^2 V_{the}^2) - \omega^4 b_1 + \omega^3 b_2 - \omega^2 b_3 + \omega b_4 - b_5 = 0. \tag{15}$$

Here

$b_1 = \omega_{pe}^2 (2k_x \upsilon_{i0} + a_3) + 2k_x \upsilon_{i0} \{\Omega^2 + k^2 V_{the}^2\} + a_2$,

$b_2 = \left[\omega_{pe}^2 (a_1 + 2k_x \upsilon_{i0} a_3) + 2k_x a_2 \upsilon_{i0} + (\Omega^2 + k^2 V_{the}^2)\{a_1 - \omega_{pi}^2 - \omega_{pd}^2\}\right]$,

$b_3 = \left[\omega_{pe}^2 a_1 a_3 + a_2(a_1 - \omega_{pi}^2) - 2\omega_{pd}^2 k_x \upsilon_{i0}(\Omega^2 + k^2 V_{the}^2) - \omega_{pd}^2 a_2\right]$,

$b_4 = -\omega_{pd}^2 (\Omega^2 a_1 + k^2 V_{the}^2 a_1 + 2k_x \upsilon_{i0} a_2)$,

$b_5 = -\omega_{pd}^2 a_1 a_2$,

together with

$a_1 = k_x^2 \upsilon_{i0}^2 - k^2 V_{thi}^2$,

$a_2 = k_y \upsilon_{e0}(\Omega^2 + k^2 V_{the}^2) + i\nu k^2 V_{the}^2$,

$a_3 = (k_y \upsilon_{e0} + i\nu)$.

Eq. (15) is the dispersion equation governing the electrostatic waves in the Hall thruster's channel.

## III. LIMITING CASES: NO DUST AND LOW TEMPERATURE

In order to check the consistency of our dispersion equation (15) with already known regimes, we discuss now its limiting cases analysed in Refs. 14, 26 and 27. The expression (15) reduces exactly to Eq. (18) of Ref. 26 in the absence of dust and also our Eq. (14) takes the following form

$$\frac{\omega_{pe}^2 \hat{\omega}}{\Omega^2(\omega - k_y v_{e0}) + \hat{\omega} k^2 V_{the}^2} + \frac{(\omega - k_x v_{i0})^2 - k^2 V_{thi}^2 - \omega_{pi}^2}{(\omega - k_x v_{i0})^2 - k^2 V_{thi}^2} = 0, \tag{16}$$

which matches with Eq. (17) of Ref. 26.

Moreover, in the absence of thermal effects ($T_i = T_e = 0$), Eq. (16) reduces to

$$\frac{\omega_{pe}^2 (\omega - k_y v_{e0} - i\nu)}{\Omega^2 (\omega - k_y v_{e0})} + \frac{(\omega - k_x v_{i0})^2 - \omega_{pi}^2}{(\omega - k_x v_{i0})^2} = 0. \tag{17}$$

The above equation reads

$$1 - \frac{\omega_{pi}^2}{(\omega - k_x v_{i0})^2} + \frac{\omega_{pe}^2}{\Omega^2} - \frac{\omega_{pe}^2}{\Omega^2} \frac{i\nu}{(\omega - k_y v_{e0})} = 0. \tag{18}$$

This relation matches with Eq. (16) of Litvak and Fisch[14]. Litvak and Fisch[14] further made use of the assumption $\omega \ll |k_y v_{e0}|$ and considered the waves propagating only along $\hat{y}$ direction, i.e. neglected $k_x$ in comparison with $\omega / v_{i0}$. Under this situation, Eq. (18) yields

$$\omega^2 = \frac{\omega_{pi}^2}{\left[1 + \frac{\omega_{pe}^2}{\Omega^2} + \frac{i\nu}{k_y v_{e0}} \frac{\omega_{pe}^2}{\Omega^2}\right]}. \tag{19}$$

viz.

$$\omega^2 = \frac{\omega_{pi}^2}{\left(\frac{\omega_{pe}^2 + \Omega^2}{\Omega^2}\right)\left[1 + \frac{i\nu \omega_{pe}^2}{k_y v_{e0}(\omega_{pe}^2 + \Omega^2)}\right]}. \tag{20}$$

Since the last term in the second square bracket of denominator in the right-hand side of Eq. (20) is small, we obtain

$$\omega \approx \pm \sqrt{\frac{\omega_{pi}^2 \Omega^2}{(\omega_{pe}^2 + \Omega^2)}} \left[1 - \frac{i\nu \omega_{pe}^2}{2k_y v_{e0}(\omega_{pe}^2 + \Omega^2)}\right]. \tag{21}$$

The corresponding real frequency is therefore simply the lower hybrid frequency, $\omega_r = \omega_{LHF}$,

$$\omega_{LHF} = \sqrt{\frac{\omega_{pi}^2 \Omega^2}{(\omega_{pe}^2 + \Omega^2)}}, \qquad (22)$$

Finally, the growth rate $-\text{Im}(\omega)$ of the resistive instability is easily expressed in the dimensionless form, normalized by the ion plasma frequency, $\gamma = -\text{Im}(\omega)/\omega_{pi}$ from Eq. (21). It is given by

$$\gamma \approx \omega_{LHF} \frac{\nu}{2k_y \upsilon_{e0}} \left[ \frac{\omega_{pe}^2}{(\omega_{pe}^2 + \Omega^2)} \right]. \qquad (23)$$

In the limit $\Omega \ll \omega_{pe}$, the growth rate reduces to

$$\gamma \approx \omega_{LHF} \frac{\nu}{2k_y \upsilon_{e0}} \qquad (24)$$

in agreement with Eq. (21) of Litvak and Fisch.[14] On the other hand, if the ionization considered in the dust-free plasma of Ref. 27 is neglected, our reduced equation (in the absence of dust) takes the form of Eq. (16) of Ref. 27.

## IV. RESULTS AND DISCUSSION

To solve the dispersion equation (15) numerically, we consider a Hall thruster plasma with parameters[7,19,26-38] $B = 100 - 250\,\text{G}$, $n_{i0} = 1 \times 10^{18}/\text{m}^3$, $n_{d0} = 1 \times 10^{14} - 5 \times 10^{14}/\text{m}^3$, $T_i = 0 - 1\,\text{eV}$, $T_e = 10 - 40\,\text{eV}$, $\upsilon_{i0} = 1 \times 10^4 - 2 \times 10^4\,\text{m/s}$, $\upsilon_{e0} = 1 \times 10^6 - 4 \times 10^6\,\text{m/s}$, $\nu = (1-7) \times 10^6/\text{s}$, $Z_d = 1 \times 10^2 - 5 \times 10^2$. For a thruster channel length $d = 4 - 10$ cm, the physically relevant wavenumbers are larger than $1/d = 10 - 25$ rad/m. The electron density is $n_{e0} = n_{i0} - Z_d n_{d0}$ thanks to quasineutrality. These sets of parameters considered for a Hall thruster reveal through the numerical calculations of Eq. (15) that the wave velocity remains always larger than the thermal velocity of the ions. Hence, the considered hydrodynamic approximation in the present calculations is a fairly good approximation for studying the influence of finite temperature in Hall thruster plasmas.[39-41]

Further, in order to investigate the variation of the growth rate of the growing wave in the presence of negatively charged dust grains with magnetic field $\vec{B}$, dust density $Z_d n_{d0}$, azimuthal wave number $k_y$, ion temperature $T_i$, electron temperature $T_e$, collision frequency $v$, we have prepared Figs. 1 – 6. In all the figures, $\gamma = -\text{Im}(\omega)/\omega_{pi}$ denotes the normalized growth rate (normalized with the ion plasma frequency $\omega_{pi}$).

In Fig. 1, we show the effect of density of dust contamination on the instability growth rate and observe that the growth rate attains slightly higher values with the larger amount of dust in the range of interest. Similar to the effect of dust density, the growth of the instability is enhanced with the enhancement of the charge and lowering of the mass of the dust grains. This is plausible as the inter-grain distance decreases with the increase of dust grain density, which increases the inter-grain repulsion due to Coulomb forces, which may also affect the dynamics of collisions and grain charging.[42–49] This results in a stronger distorting force that unsettles the equilibrium and enhances the growth of the instability. Similarly, the increase of the charge on the dust grains results in the increase of distorting force due to larger Coulomb force, which unsettles the equilibrium of the plasma.[32] It can be noticed that the influence of mass of dust particle is very small: changing it by five orders of magnitude only changes the growth rate by a few percents. As in a real thruster there is a distribution of mass and that this mass distribution is hardly known, it is a good point that will simplify comparison to experiment. In the range of values scanned on Fig. 1, neither does the product $Z_d n_{d0}$ change the growth rate significantly (see also Fig. 6, later).

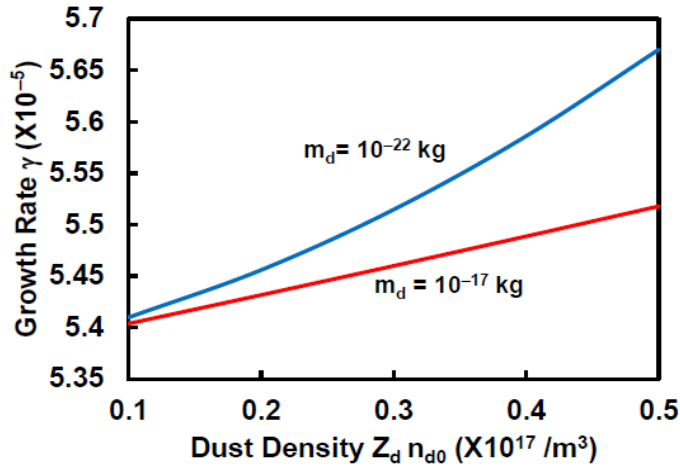

Fig. 1. Variation of growth rate (normalized to ion plasma frequency) $\gamma$ with dust density in the plasma with Xe ions ($M$ = 131 amu), when $T_e = 25\,\text{eV}$, $T_i = 1\,\text{eV}$, $n_{i0} = 10^{18}/\text{m}^3$, $B = 0.015\,\text{T}$, $k_y = 400/\text{m}$, $k_x = 20/\text{m}$, $\upsilon_{i0} = 10^4\,\text{m/s}$, $\upsilon_{e0} = 10^6\,\text{m/s}$, $Y_e = Y_i = 2$, $V_{thi} = 1.2\times 10^3\,\text{m/s}$, $V_{the} = 3\times 10^6\,\text{m/s}$, $\Omega = 4.4\times 10^9/\text{s}$ and $v = 10^6/\text{s}$.

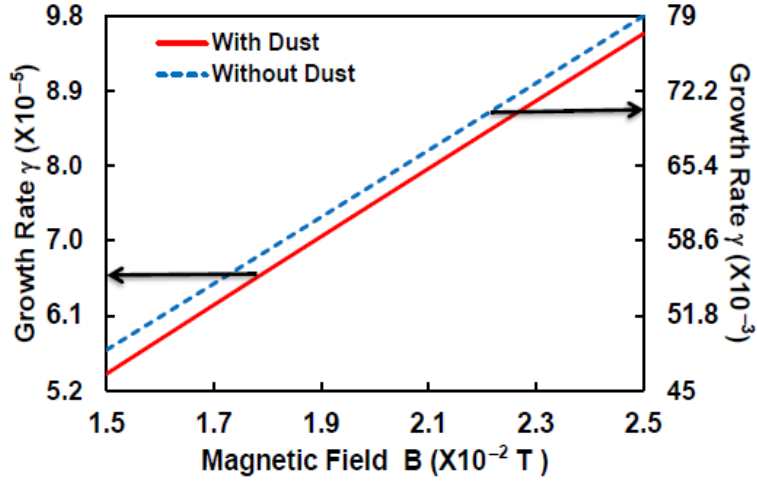

Fig. 2. Dependence of growth rate $\gamma$ on magnetic field in the presence of dust (solid line) and in the absence of dust (dashed line), when $m_d = 10^{-17}$ kg, $Z_d n_{d0} = 1 \times 10^{16} / m^3$ and the other parameters are the same as in Fig.1. Caution: each curve has its own scale (follow the arrows).

Figure 2 shows the dependence of the instability growth rate $\gamma$ on the magnetic field with and without dust. One also observes from the figure that the instability growth rate increases in the presence of stronger magnetic field. This variation is in agreement with the experimental and theoretical results of Wei *et al.*[31] The larger growth rate can be attributed to the nonuniformity of the velocity in the presence of a stronger magnetic field. Moreover, the presence of the magnetic field confines the electrons' trajectories and increases their residence time in the channel, which will enhance the oscillation frequency. This will result in a more unstable situation and hence the instability grows faster.

A most striking result is that the magnitude of growth rate in the presence of dust particles is about $10^{-3}$ times smaller than the growth rate without dust particles[26], which is shown on the secondary axis of the figure.

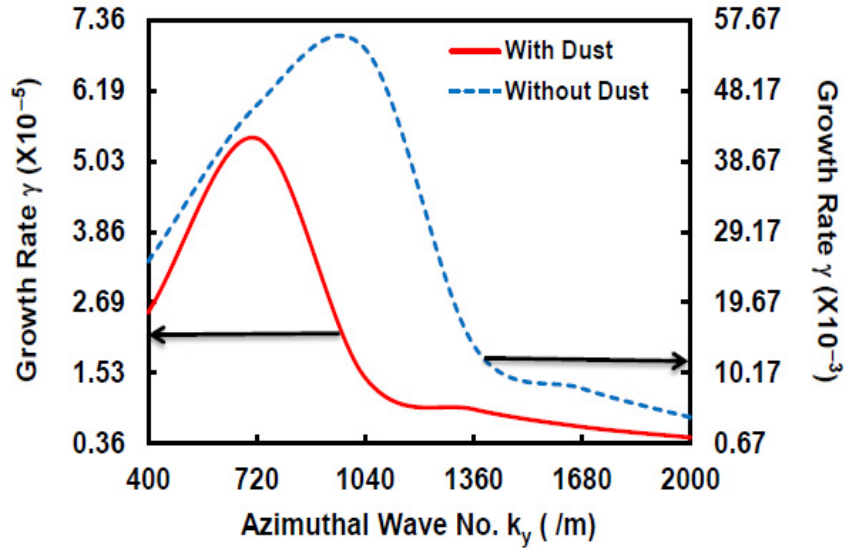

Fig. 3. Dependence of growth rate $\gamma$ on azimuthal wave number ($k_y$) in the presence of dust (solid line) and in the absence of dust (dashed line), when $m_d = 10^{-17}$ kg, $Z_d n_{d0} = 1 \times 10^{16} / m^3$ and the other parameters are the same as in Fig.1.

Fig. 3 shows the effect of azimuthal wave number on the growth rate of the instability. It is found that the growth rate first increases for the oscillations of higher azimuthal wave number, i.e. oscillations of smaller wavelength are more susceptible to the instability; then, after attaining a peak, the growth rate goes down with the higher values of the wave number. The position of the peak of the growth rate in the presence of the dust particles is shifted (noticeably, but within the same order of magnitude) from the peak attained by the growth rate when the dust particles are absent in the plasma. Clearly the difference is caused by the heavy dust particles that also oscillate and moderate the plasma dynamics.

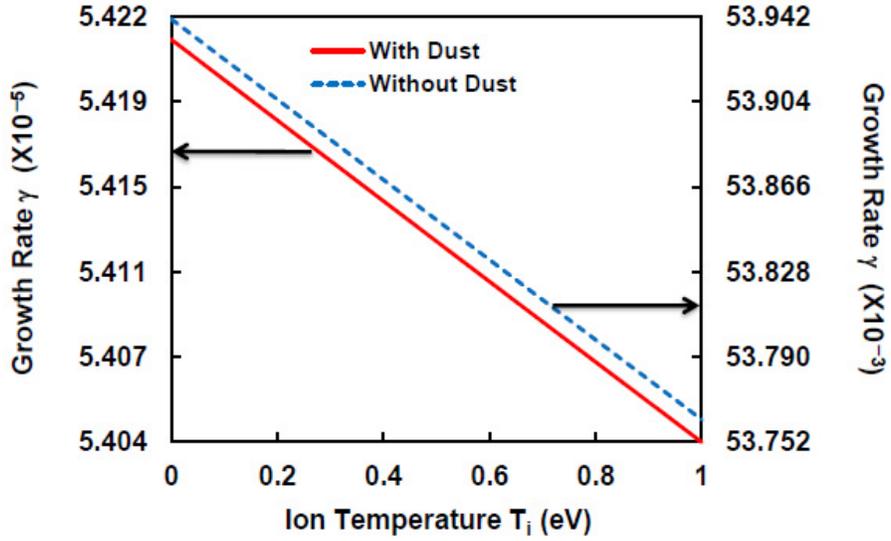

Fig. 4. Weak dependence of growth rate $\gamma$ on ion temperature with dust grains (solid line) and without dust grains (dashed line), when $m_d = 10^{-17}$ kg, $Z_d n_{d0} = 1 \times 10^{16} / m^3$ and the other parameters are the same as in Fig. 1.

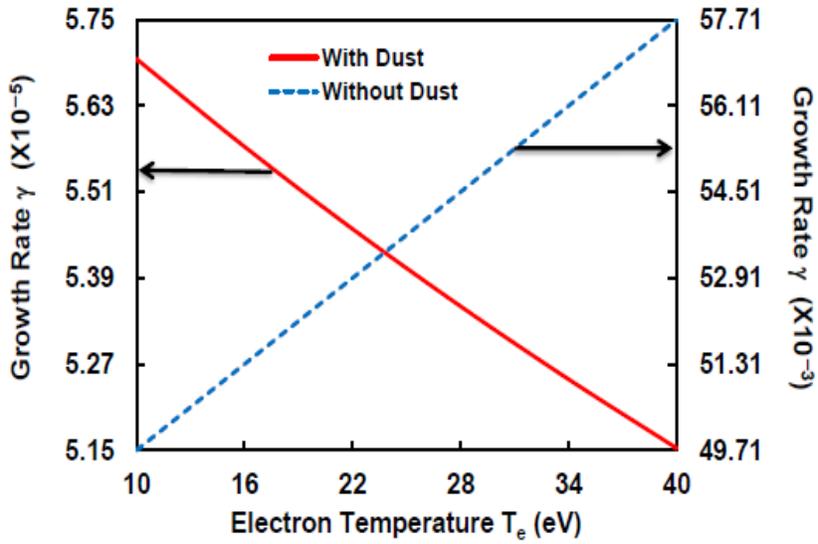

Fig. 5. Dependence of growth rate $\gamma$ on electron temperature with dust grains (solid line) and without dust grains (dashed line), when $m_d = 10^{-17}$ kg, $Z_d n_{d0} = 1 \times 10^{16} / m^3$ and other parameters are the same as in Fig. 1.

The effect of ion and electron temperatures on the growth rate of the instability is shown in Fig. 4 and Fig. 5, respectively. The higher temperature of either species in the present case of contaminated plasma is found to reduce the instability growth rate, the effect of ion temperature on growth rate being very weak (0.3%) whereas the effect of electron temperature is more sensitive (10%) on the typical interval of variation of those parameters relevant for Hall thrusters. However, the opposite effect of electron temperature on the growth rate was observed in dust-free plasma.[26] This could also be due to the presence of dust particles which oscillate and curb the excitation of the wave and instability.

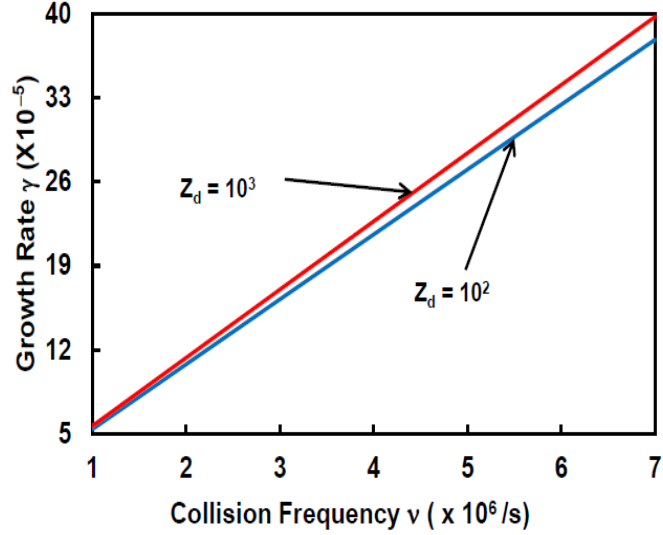

Fig. 6. Variation of growth rate $\gamma$ with collision frequency, when $m_d = 10^{-17}$ kg, $n_{d0} = 1 \times 10^{14} / m^3$ and the other parameters are the same as in Fig. 1.

We show the variation of the growth rate with collision frequency in Fig. 6, where the growth rate is found to increase significantly with the higher values of the collision frequency $v$. The growth rate of the instability appears to be directly proportional to the collision frequency, as was found in Ref. 27. This is possible as the resistive coupling is enhanced in the presence of more collisions. However, the simulation studies of resistive instability by Fernandez et al.[33] showed the growth rate to be directly proportional to the square root of the collision frequency. Further, in our case, the impact of collisions on the instability growth becomes more pronounced for the case of dust grains having higher charge. This is due to the fact that the plasma dynamics is sensitively altered in the presence of dust particles having higher charge.

Finally, we investigate the variation of the wave real frequency $\omega_r$ with the magnetic field $B$ and the dust density $Z_d n_{d0}$ for studying the propagation characteristics. We find actually a phase velocity of the same order of magnitude as the electron drift velocity, which is coherent with experimental observations.[8, 19] Figure 7 shows both cases $k_x \ll k_y$ and $k_x = k_y$. The effect of magnetic field is found to increase the real frequency, and hence the phase velocity of the wave. Moreover, Fig. 7(b) shows that, for a given magnetic field and given azimuthal wavenumber $k_y$, the instability propagates at higher speed when the axial component becomes significant and the condition $k_x = k_y$ is satisfied. In view of the higher growth rate and larger phase velocity of the instability in the situation of significant axial wave number $k_x$, we can say that the instability propagates at a larger velocity in the case of its faster growth. Similar results are obtained for the situation of plasma having higher dust contamination and the lighter particles, which is evident from Fig. 8. The enhancement of the real frequency with the axial and azimuthal wave numbers in Figs. 7 and 8 is consistent with the observation made by Fernandez et al.[33].

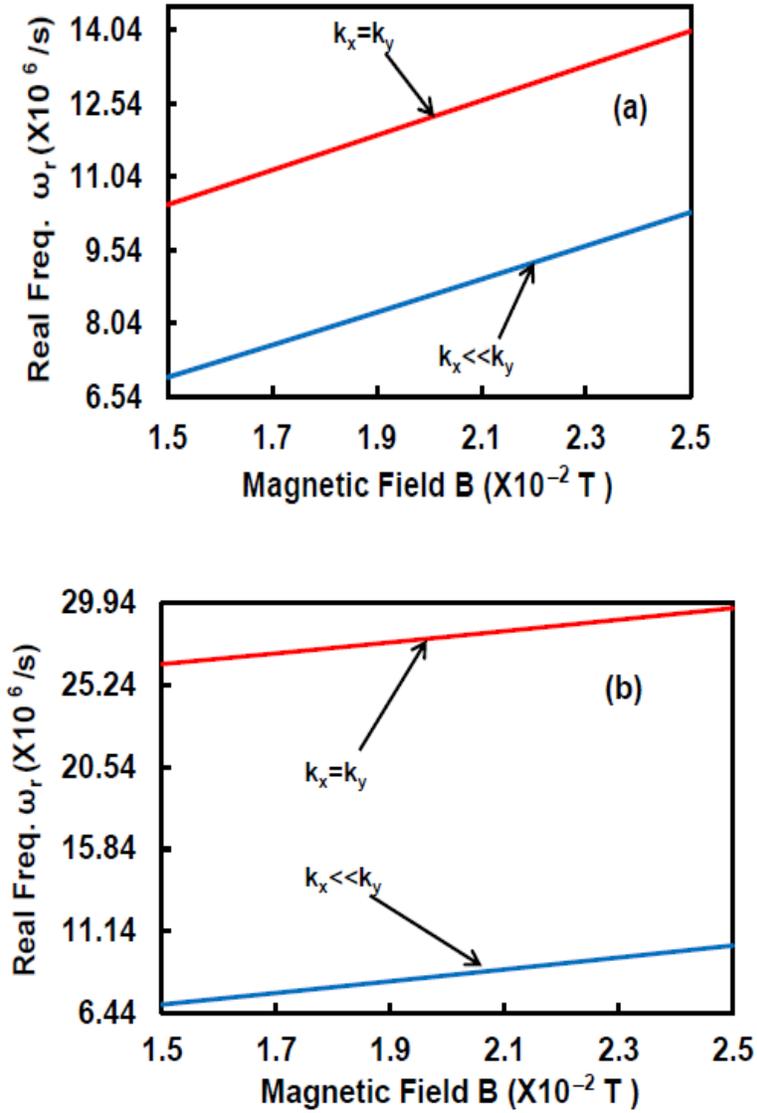

Fig. 7. Variation of real frequency $\omega_r$ with magnetic field, when $k_x \ll k_y$ ($k_x = 20$/m and $k_y = 400$/m), $m_d = 10^{-17}$ kg, $Z_d n_{d0} = 1\times 10^{16}/m^3$ and the other parameters are the same as in Fig. 1. The equal wave number case is considered with $k_x = k_y = 100$/m in (a), whereas $k_x = k_y = 400$/m in (b).

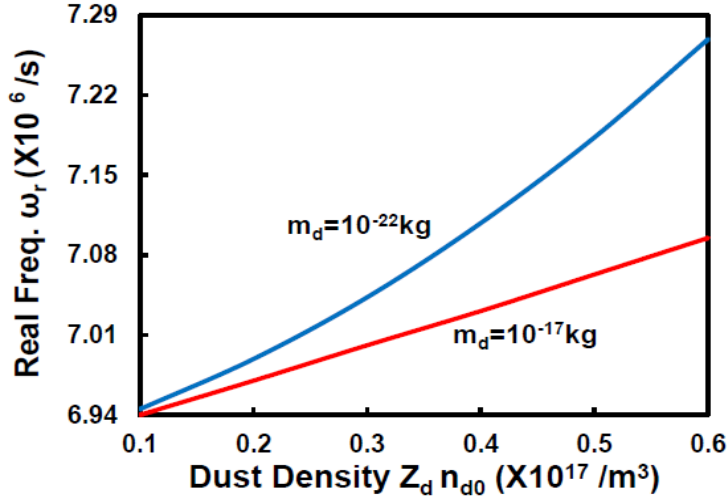

Fig. 8. Variation of real frequency $\omega_r$ with dust density, when the parameters are the same as in Fig. 1.

Through the results from Figs. 1 – 8, we examined the resistive instability in a Hall thruster plasma. In the thruster channel, the azimuthally propagating waves become unstable due to the resistive coupling of disturbances to the electrons' $\vec{E}\times\vec{B}$ drift in the presence of collisions. In general, such types of disturbances have both axial and azimuthal components, and can propagate with an angle α in the Hall thrusters, but the situation is different in the case of usual collisional magnetized dusty plasmas for the instabilities[48] and collective effects[49] in complex plasma.

**V. CONCLUSIONS**

In this paper, we presented an analytical approach to study the instability in dusty Hall plasmas where an azimuthally propagating wave (wave number $k_y$) grows due to the coupling of electron drift with the oscillations in the presence of collisions. Since an axial component of wave vector $k_x$ also appears in such instabilities, we considered two cases, one with $k_x \ll k_y$ and one with $k_x = k_y$, and found that, for a fixed azimuthal component $k_y$, the instability grows faster when the axial wavenumber $k_x$ is larger.

The main focus in this work was on the impact of dust on the growth of this instability. The magnitudes of growth rates in the presence of dust and in its absence are found to be quite different. On the other hand, the growth rate decreases with the higher thermal motions. The instability grows even faster in the presence of dust when the dust carries lower mass and larger charge. As expected, the higher collision frequency enhances the growth rate due to stronger coupling of the electrons drift with the oscillations. This effect becomes more significant in the presence of higher charge of the dust. In addition, the magnetic field was found to enhance the growth of instability.

## ACKNOWLEDGEMENT

The authors acknowledge the financial assistance given by Indo French Centre for the Promotion of Advanced Research (IFCPAR/CEFIPRA), New Delhi. Assistance from L. Couëdel was helpful to YE.